\documentstyle[preprint,epsfig,aps]{revtex}
\begin{document}
\draft
\preprint{\begin{tabular}{r}
{\bf hep-ph/0008073}\\
KIAS-P00036
\end{tabular}}

\title{CP violating supersymmetric contributions to the electroweak 
$\rho$ parameter}
\author{Sin Kyu Kang$^1$\footnote{skkang@ns.kias.re.kr} and Jung Dae Kim$^{1,2}$
 \footnote{jdkim@theory.yonsei.ac.kr}}
\address{$^1$ School of Physics, Korea Institute for Advanced Study, Seoul
130-012, Korea \\
$^2$Physics Department and IPAP, Yonsei University, Seoul 120-749, Korea 
}


\maketitle     

\begin{abstract}
\noindent
Effects of CP violation on the supersymmetric electroweak
correction to the $\rho$ parameter are  investigated.
To avoid the EDM constraints, we require that arg$(\mu)<10^{-2}$ and
the non-universal trilinear couplings $A_f=(0,0,A_0)$ and also assume that 
gluinos are heavier than 400 GeV. 
The CP phase $\phi_t=$ arg($A_0$) leads to large enhancement of the 
relative mass splittings between $\tilde{t}_2$ and $\tilde{b}_L(\tilde{t}_1)$,
which in turn reduces the one-loop contribution of the stop and sbottom to 
$\Delta \rho$. For small $\tan \beta$, such a CP violating effect is prominent.
We also study how much the two-loop gluon and gluino contributions
are affected by the CP phase. Possible contributions to the $\rho$ parameter
arising from the Higgs sector with CP violation are discussed. 
\end{abstract}


The minimal supersymmetric standard model (MSSM) is the best motivated 
extensions of the standard model (SM).
As is well known, in the MSSM, there are many new sources of CP violation
beyond the Cabibbo-Kobayashi-Maskawa (CKM) phase, arizing from the complex 
soft supersymmetry (SUSY) breaking terms, i.e., 
the Majorana gaugino masses $M_i$, the trilinear scalar $A$ terms, 
the bilinear scalar $B$ terms, as well as from the $\mu$ parameter which
is the bilinear mixing of the two Higgs chiral superfields in the 
superpotential.
While those supersymmetric CP violating (CPV) phases could give large 
contributions to the neutron/electron electric dipole moments (EDMs),
they are strongly constrained by the current experimental measurements of
the neutron/electron EDMs, which for the neutron is
$d_n < 1.1\times 10^{-25} e$ cm \cite{edm1} and for the electron is 
$d_e < 4.3 \times 10^{-27} e$ cm \cite{edm2}.
To resolve this problem, several possibilities have been suggested. 
The first one is to make all phases small ($O(10^{-2}$)) \cite{edm3}.
But, such small phases would require a significant amount of fine-tuning 
and are thus undesirable.
The second is to take the SUSY spectrum heavy in the several TeV range which
lies outside the reach of the accelerators \cite{edm4}. Another option
suggested recently is to use the cancellations among the various contributions
to the neutron/electron EDMs, allowing for large CPV phases
and a supersymmetric spectrum that is still within the reach of the
accelerators \cite{edm5}. 
Finally, an interesting possibility is to take a slightly
non-universal scenario for the trilinear couplings $A_f$ \cite{edm6,edm7}.
As shown in Ref. \cite{edm7}, requiring that arg($\mu)<10^{-2}$ and
$A_f=(0,0,A_0)$, and assuming that gluinos are heavier than
about 400 GeV \cite{edm5}, one may significantly reduce the size of the
neutron/electron EDMs due to Weinberg's three-gluon operator \cite{weinberg}
well below the current experimental limit.

Another CP violation associated with the Higgs boson sector
in the MSSM can come from the one-loop corrections of the MSSM Higgs 
potential through soft CPV Yukawa interactions involving squarks, i.e.,
$A$ terms. 
As shown in Ref.\cite{higgscp1}, an immediate consequence of 
Higgs sector CP violation in the MSSM is the generation of mixing mass terms
between the CP even and CP odd Higgs fields. 
Thanks to those scalar-pseudoscalar mixing mass terms, the small tree-level
mass difference between the heaviest Higgs boson and the CP-odd scalar may
be lifted considerably.
The phenomenological consequences of Higgs sector CP violation have been
studied \cite{higgscp1,higgscp2}.

A possible way to probe SUSY is to search for the virtual effects of 
SUSY particles which would be sizable enough to be detected in the present
experiments. In most experimentally accessible processes, the heavy
SUSY particles decouple from the low energy electroweak observables. 
However, similar to the SM, if there happens large "custodial" $SU(2)_V$ 
breaking, the electroweak observables which are parameterized
by the $\rho$ parameter \cite{rho},
defined as the ratio of the strengths of neutral and charged currents 
at vanishing momentum transfer, can severely be affected. 
In the MSSM, the main potential source of $SU(2)_V$ breaking is
the splitting between the stop and sbottom masses. In addition, there
can be large splitting within a supersymmetric multiplet, which also
affects the $\rho$ parameter.
From the numerical analysis, it is well
known that the leading contribution of the squark loops, in particular
stop and sbottom loops, to the $\rho$ parameter can reach
at the level of a few $10^{-3}$ which lies within the range of the
experimental sensitivity \cite{rhosusy1}.
There are also small additional contributions coming from the 
exchange of the additional Higgs bosons, the charginos and neutralinos. 
Some discussions of supersymmetric contributions to the $\rho$ parameter
already exist in the literature \cite{rhosusy1,rhosusy2}. 
Moreover, higher order corrections have also been studied so as to
treat the SUSY loop contributions to the electroweak observables 
at the same level of accuracy as the SM contribution \cite{rhosusy3}.
However, so far, the analyses have been done within the context of the CP
conserving (CPC) MSSM.
Since new CPV phases may affect the SUSY spectrum and generate 
additional couplings which do not appear in the CPC MSSM, 
it would be quite interesting to study the effects of the CPV SUSY phases
on the electroweak observables via the $\rho$ parameter.

In this respect, the purpose of this letter is to examine how much the
$\rho$ parameter can be affected by the non-vanishing CP phases in the MSSM.
In this analysis, we will impose the universal conditions on the
soft SUSY breaking terms which provide three complex parameters, 
i.e., universal gaugino mass $m_{1/2}$, the trilinear soft term $A_0$ and
the bilinear soft term $B_0$. In addition, there is a complex mass
parameter $\mu$.
However, not all phases of the four complex parameters are
physical \cite{cpv1}. It is always possible to make a phase 
transformation on the gaugino fields so as to make $m_{1/2}$ real.
From minimization conditions on the Higgs potential, one can make the phase
of $B_0$ equal to the phase difference of the two Higgs doublets in the MSSM.
As a result, there are only two independent CPV phases in the
MSSM with universal soft SUSY breaking terms, which are usually chosen
to be arg($A_0$) and arg($\mu$). To satisfy the EDM constraints, we will
simply take
arg($\mu $) to be zero and the trilinear coupling of the Higgs bosons to the
squarks of the first and second generation to be much smaller than the one of
the third generation, as mentioned above,  and assume that the gluino mass to 
be larger than 400 GeV.
Thanks to this non-vanishing phase arg($A_0$), as will be shown later, 
the stop spectrum and their mixing angle are changed,
which lead in turn to the shift of the contributions to $\Delta \rho$.
Also, since the neutral Higgs boson mass eigenstates in the CPV MSSM  
are mixtures of CP even and CP odd states, new gauge-Higgs couplings 
are induced at the tree level through those mixings. These couplings can
generate the additional radiative corrections
to the gauge boson self-energies, and may thus affect
the $\rho$ parameter.
However, the contributions coming from the gaugino and higgsino exchanges 
will not be affected by the non-vanishing CPV phase arg($A_0$) in the leading 
order because they are affected by only the CPV phase arg($\mu$) which is 
neglected in this analysis. Thus, we shall not consider those contributions.
In addition, we shall study how much the dominant two-loop contributions via
gluon and gluino exchanges can be shifted by the effect of the CPV phases.
The numerical calculation will be done with the help of the program package FEYNHIGGS
\cite{feynhiggs} which is modified so as to calculate the CPV SUSY contributions
to the $\rho$ parameter.

The $\rho$ parameter is represented by
\begin{equation}
\rho = \frac{1}{1-\Delta \rho}, ~~~~~\Delta \rho = \frac{\Pi_{ZZ}(0)}{M^2_Z}
-\frac{\Pi_{WW}(0)}{M^2_W},
\end{equation}
where $\Pi_{WW(ZZ)}(0)$ denotes the transverse parts of the $W/Z$-boson
self-energies at zero momentum transfer.
In the SM, the main contribution comes from the large splitting between the
top quark and the bottom quark masses and is given by
\begin{equation}
\Delta \rho^{SM}_0=\frac{N_cG_F}{8\sqrt{2}\pi^2}F_0(m^2_t, m_b^2),
\end{equation}
where $N_c$ is the color factor and the function $F_0$ is given by
\begin{equation}
F_0(x,y)=x + y - \frac{2xy}{x-y}\ln \frac{x}{y}.
\end{equation}
In the limit of large mass splitting it becomes proportional to the heavy
quark mass squared, i.e., $F_0(m^2_t, m_b^2) \simeq m_t^2$.
Including QCD corrections, the SM contribution of $\Delta \rho$ is
$\Delta \rho^{SM}_1=-\Delta \rho_0^{SM}\left(\frac{2\alpha_s}{3 \pi}\right)
(1+\pi^2/3)$ \cite{qcd1}.
For $\alpha_s(M_Z)\simeq 0.12$, the QCD correction reduces the one-loop result
by about $10\%$ and shifts $m_t$ upwards by about $10$ GeV \cite{qcd2}.

{\bf (A) Squark contributions}:
Since the scalar partners of the light quarks are expected to be almost 
degenerate in most supersymmetric theories, their contributions to $\Delta \rho$
are negligible and thus only third generation will contribute. 
The stop mass matrix is given by
\begin{eqnarray}
M^2_{\tilde{t}}=\pmatrix{m^2_{\tilde{t}_L}+m^2_t+D_L & m_t m^{\ast}_{LR} \cr
       m_tm_{LR} & m_{\tilde{t}_R}^2+m_t^2+D_R}
\end{eqnarray}
where $m_{LR}=A_0-\mu^{\ast}/\tan \beta$, $D_L=\left(\frac{1}{2}-\frac{2}{3}
\sin^2\theta_W\right)\cos2\beta M^2_Z$ and $D_R=\frac{2}{3}\sin^2\theta_W
\cos2\beta M^2_Z$. The stop fields $(\tilde{t}_L, \tilde{t}_R)$ are
linear combinations of the mass eigenstates $\tilde{t}_{1,2}$ so that
$\tilde{t}_L=\cos\theta_{\tilde{t}}\tilde{t}_1-\sin\theta_{\tilde{t}}
e^{-i\varphi_t} \tilde{t}_2, ~
\tilde{t}_R=\sin\theta_{\tilde{t}}e^{i\varphi_t}\tilde{t}_1+\cos\theta_{\tilde{t}}
\tilde{t}_2$, 
where
\begin{equation}
\tan 2\theta_{\tilde{t}} = \frac{2m_t|m_{LR}|}
{m^2_{\tilde{t}_L}- m^2_{\tilde{t}_R}+D_L-D_R},
\end{equation}
and the mass eigenvalues of stop fields are given by
\begin{eqnarray}
m^2_{\tilde{t}_{1,2}} &=& \frac{1}{2}\left[m^2_{\tilde{t}_L}+ m^2_{\tilde{t}_R}
+2m^2_t+D_L+D_R \right. \nonumber \\
& &\left. \mp \left([m^2_{\tilde{t}_L}- m^2_{\tilde{t}_R}+D_L-D_R]^2
+4m^2_t|m_{LR}|^2 \right)^{1/2}\right],  \\
|m_{LR}| &=& |A_0-\mu^{\ast}/\tan\beta| \nonumber \\
         &=& \sqrt{|A_0|^2+|\mu|^2/\tan^2\beta -2
|A_0||\mu|/\tan\beta\cos\phi_t},
\end{eqnarray}
where $\phi_t\equiv \mbox{arg}(A_0)$.
As expected, the non-vanishing phase $\phi_t$  will
change the stop mass spectrum which can in turn lead to the shift of $\Delta
\rho$.
Neglecting the mixing in $\tilde{b}$ sector, the contribution of $(\tilde{t},
\tilde{b})$ to $\Delta \rho$ is presented at one-loop order by:
\begin{eqnarray}
\Delta \rho_{\tilde{t}\tilde{b}} &=& \frac{3G_F}{8\sqrt{2}\pi^2}
[-\sin^2\theta_{\tilde{t}}\cos^2\theta_{\tilde{t}}F_0(m^2_{\tilde{t_1}},
m^2_{\tilde{t_2}}) \nonumber \\
&+&\cos^2\theta_{\tilde{t}}F_0(m^2_{\tilde{t_1}}, m^2_{\tilde{b_L}})
+\sin^2\theta_{\tilde{t}}F_0(m^2_{\tilde{t_2}}, m^2_{\tilde{b_L}})],
\end{eqnarray}

Before examining how much the value of $\Delta \rho_{\tilde{t}\tilde{b}}$
can be affected by the phase $\phi_t$,
let us first investigate the mass splittings between
the two squark mass eigenstates which may indicate to what extent the
contribution of $(\tilde{t}, \tilde{b})$ to $\Delta \rho$ is deviated.
In Fig. 1 a (b), we present numerical
estimation of the relative mass splittings between the two squark mass 
eigenstates, $|m_{\tilde{t}_1}^2-m_{\tilde{t}_2}^2|/m^2_{\tilde{q}}$ and 
$|m_{\tilde{t}_2}^2-m_{\tilde{b}_L}^2|/m^2_{\tilde{q}}$, as a 
function of the phase $\phi_t$, for $\tan\beta=1.6 (20), m_{\tilde{q}}=200$
GeV, $|A_0|=200$ GeV and $|\mu|=100$ GeV. Here we take $0\leq \phi_t \leq \pi$. 
As the phase $\phi_t$ increases, the both relative mass splittings increase.
The effect of the maximal CP violation corresponding to $\phi_t=\pi$
may lead to the enhancement of the relative mass splittings by about
$90\%(40\%)$ for $|m_{\tilde{t}_1}^2-m_{\tilde{t}_2}^2|/m^2_{\tilde{q}}$ 
($|m_{\tilde{t}_2}^2-m_{\tilde{b}_L}^2|/m^2_{\tilde{q}}$) compared to the
CPC case.
This is attributed  to the fact that the absolute value of $m_{LR}$
monotonically increases with the increasing phase
$\phi_t$, which makes the mass splitting between $\tilde{t}_1 (\tilde{b}_L)$ 
and $\tilde{t}_2$ larger, and the mixing angle $\theta_{\tilde{t}}$ closer to
$\pi/4$.  Note that the mixing angle $\theta_{\tilde{t}}$ close to
$\pi/4$ are naturally obtained because the off-diagonal elements
of the stop mass matrix may be larger than the difference of the
diagonal entries. As one can see from Fig. 1 a, 
in the case of small $\tan\beta$, the splitting
$|m_{\tilde{t}_2}^2-m_{\tilde{b}_L}^2|/m^2_{\tilde{q}}$ is larger than
$|m_{\tilde{t}_1}^2-m_{\tilde{t}_2}^2|/m^2_{\tilde{q}}$ for $\phi_t=0 $
(CPC case), 
whereas the latter becomes larger than the former for large $\phi_t$. 
Thus, the splitting originating from the left-right
mixing of stops may be more important for the CPV case with 
large phase. In the case of large $\tan\beta$, 
$|m_{\tilde{t}_1}^2-m_{\tilde{t}_2}^2|/m^2_{\tilde{q}}$ is larger than
$|m_{\tilde{t}_2}^2-m_{\tilde{b}_L}^2|/m^2_{\tilde{q}}$ for all $\phi_t$,
and the mass splittings are weakly  dependent on $\phi_t$ 
since the term concerned with the phase $\phi_t$ in $|m_{LR}|$ is 
suppressed by the large $\tan \beta$. 
We also observed that the mass splitting $|m_{\tilde{t}_1}^2
-m_{\tilde{b}_L}|/m^2_{\tilde{q}}$ is much smaller than the other two. 
For the same input values of the SUSY parameters,
the dependence of $\Delta \rho_{\tilde{t}\tilde{b}}$ on the phase $\phi_t$
is shown as solid lines in Fig. 2. 
Although the CPV case provides larger mass splitting than the CPC case,
$\Delta \rho_{\tilde{t}\tilde{b}}$ is monotonically decreased with the
increasing $\phi_t$.
The effect of the CPV phase reduces the value of 
$\Delta \rho_{\tilde{t}\tilde{b}}$ for the CPC case by about up to $35\%$.
The reason is that as the phase increases, the mixing angle 
$\theta_{\tilde{t}}$ gets close to the maximal mixing and 
$F_0(m^2_{\tilde{t}_1},
m^2_{\tilde{t}_2})$ increases faster than $F_0(m^2_{\tilde{t}_2},
m^2_{\tilde{b}_L})$ in Eq. (8) so that they lead to destructive contribution
to $\Delta \rho_{\tilde{t}\tilde{b}}$. 
As $|A_0|$ increases, $\Delta \rho_{\tilde{t}\tilde{b}}$ tends to decrease, 
whereas it tends to increase with the increasing $|\mu|$. 
We have observed that the effect of the CPV phase is diminished 
as long as $m_{\tilde{q}}$ becomes larger
than $|A_0|$ and $|\mu|$. This is because $m_{\tilde{t}_1}\sim 
m_{\tilde{t}_2}\sim m_{\tilde{b}_L}$ in the limit of large 
$m_{\tilde{q}}$. 

The two-loop QCD correction to $\Delta \rho_{\tilde{t}\tilde{b}}$
is given by Eq.(8) in Ref.\cite{rhosusy3}, and
the two-loop contribution mediated by gluino exchange is also represented
in Ref.\cite{rhosusy3}.  Those gluon and gluino contributions
add up to about $30\%$ of the one-loop value in the CPC case.

In Fig. 3, the dependence of the dominant two-loop QCD correction and gluino 
contribution to the $\rho$ parameter on the phase $\phi_t$ is 
plotted by the dashed line and dotted line, respectively,
for $\tan \beta=1.6$ (a) and $\tan \beta=20$ (b).
As the phase $\phi_t$ increases, the two-loop gluonic SUSY contribution 
decreases, whereas the gluino contribution increases.
The reason that the gluino contribution increases with the increasing phase
$\phi_t$ is that its contribution depends inversely on the mass splittings
between $\tilde{t}_1(\tilde{b}_L)$ and $\tilde{t}_2$.
In particular, for small $\tan \beta$, 
the value of the gluino contribution becomes larger than that of the gluonic
SUSY contribution at large $\phi_t$.

{\bf (B) Higgs sector contributions}:
There are also possible contributions arising from 
the Higgs sector to the $\rho$ parameter.
In the CPC case, it is known \cite{rhosusy1} that the Higgs boson masses and 
couplings 
to gauge bosons are related in such a way that they lead to large cancellations
in their contributions to the $\rho$ parameter, which is at most of order
of $10^{-4}$.
In the decoupling limit where the heavy neutral CP-even Higgs, CP-odd 
Higgs and the charged Higgs bosons are nearly degenerate and their
couplings to gauge bosons tend to zero, while the lightest CP-even Higgs
boson reaches its maximal mass value and has almost the same properties
as the SM Higgs boson. Then, the contribution of the Higgs sector of the
MSSM to the $\rho$ parameter is practically the same as in the SM.
However, while the contribution of the SM Higgs sector gives rise to
logarithmic $\log M_h/M_Z$ in the limit of large Higgs mass which
may reach about $10^{-3}$ order, the MSSM contribution reaches at most
$10^{-4}$ due to the upper bound on the mass of the lightest Higgs boson.

Now, let us take into account the contributions coming from the neutral
Higgs sector with CP violation. The decomposition of the two
Higgs doublets is given by
\begin{eqnarray}
\Phi_1=\pmatrix{\phi_1^+ \cr \frac{1}{\sqrt{2}}(v_1+\phi_1+ia_1)},~~~
\Phi_2=e^{i\theta}\pmatrix{\phi_2^+ \cr \frac{1}{\sqrt{2}}(v_2+\phi_2+ia_2)},
\end{eqnarray}
where $v_1$ and $v_2$ are the vacuum expectation values and $\theta$ is
their relative phase. 
As shown in Ref. \cite{susyhiggs}, the Higgs quartic couplings receive 
significant radiative corrections from  $A_f$ terms.
The vacuum expectation values $v_1,v_2$ and the phase $\theta$ are determined
from the minimization conditions on the Higgs potential of the MSSM. 
The CP odd fields are rotated to
$a_1 = \cos \beta G^0 -\sin \beta a$ and $a_2=
                         \sin \beta G^0 + \cos \beta a$
so that the Higgs potential shows up a flat direction with respect to the
Goldstone field $G^0$.
The non-vanishing CPV phases in $A_t$ and $A_b$ can lead to the
non-vanishing $\theta$, for  which the scalar-pseudoscalar  mixing mass
terms are generated and the neutral Higgs boson mass matrix in the
weak basis $(G^0, a, \phi_1, \phi_2)$ can be written by
\begin{eqnarray}
M^2_0 = \pmatrix{M^2_P & M^2_{PS} \cr M^2_{SP} & M^2_S},
\end{eqnarray}
where $M_P^2$ and $M^2_S$ are $2\times 2$ matrix for
CPC parts in the basis $(G^0, a)$ and $(\phi_1, \phi_2)$, 
respectively, and $M^2_{PS}=(M^2_{SP})^T$ describes the CPV mixings
between $(G^0, a)$ and $(\phi_1, \phi_2)$.
From the tadpole condition, the Goldstone field $G^0$ does not mix with
the other neutral fields and thus becomes massless.
Then, the neutral Higgs mass matrix $M^2_0$ reduces to a $3\times 3$ matrix
$M^2$, which is spanned in the basis $(a, \phi_1, \phi_2)$.
The Higgs boson mass matrix $M^2$ can be diagonalized with the help of
an orthogonal rotation matrix $U$:
$U^{T} M^2 U = diag (M^2_{H_3}, M^2_{H_2}, M^2_{H_1})$.

Then, the contributions of the Higgs sector with CP violation to the 
$\rho$ parameter can be given by
\begin{eqnarray}
\Delta \rho_H &=& \frac{3\alpha}{16\pi c^2_W}
\sum^3_{i=1} U^2_{1i}\Delta \rho_H^{SM}(M^2_{H_i})
 + \frac{\alpha}{16\pi s^2_W M^2_W}  \nonumber \\
&\times & \left( \sum^{3}_{i=1}(1-U^2_{1i})F_0(M^2_{H_i},M^2_{H_{c}}) 
- \frac{1}{2} \sum^{3}_{i,j}(U_{1i}U_{2j}-U_{2i}U_{1j})^2
   F_0(M^2_{H_i},M^2_{H_{j}})\right) 
\end{eqnarray}
where $M_{H_c}$ denotes the charged Higgs boson mass and
\begin{eqnarray}
\Delta \rho_H^{SM}(M^2) &=&\frac{\alpha}{16\pi s^2_W M^2_W}
[F_0(M^2, M^2_W)-F_0(M^2, M^2_Z)] \nonumber \\
&+& \frac{\alpha}{16\pi s^2_W} \left[\frac{M^2}{M^2-M^2_W}\log 
\frac{M^2}{M^2_W}
- \frac{1}{c^2_W}\frac{M^2}{M^2-M^2_Z}\log \frac{M^2}{M^2_Z}\right]
\end{eqnarray} 
corresponds to the SM Higgs boson contribution to $\Delta \rho$.

In Fig. 3, we present the dependence of
the Higgs boson sector contribution to $\Delta \rho$ on the phase $\phi_t$ 
(solid lines). Similar to
the one-loop contribution of stop quark to $\Delta \rho$, the one-loop
contribution of Higgs sector is decreased with the increasing phase $\phi_t$.
This is mainly because the relative mass splittings among three neutral
Higgs bosons decrease as the phase $\phi_t$ increases as shown in Fig. 4.
In the case of small $\tan \beta$, $|M^2_{3}-M^2_{2}|/M^2_{H_c}$ (solid line)
is larger than  $|M^2_{2}-M^2_{1}|/M^2_{H_c}$ (dotted line).
For only small $\tan \beta$, the contribution of the Higgs sector is larger
than the two-loop contributions mediated by the gluon and
gluino exchanges, and those three contributions are almost the same at 
$\phi_t=\pi$, which increase the one-loop contribution of stop and sbottom 
by about $30 - 35\%$.
For large $\tan \beta$, the Higgs sector contribution to $\Delta \rho$
is negligibly small.

In summary, effects of CP violation on the supersymmetric electroweak
correction to the $\rho$ parameter have been investigated.
To avoid the EDM constraints, we have required negligibly small arg$(\mu)$
and the non-universal couplings $A_f=(0,0,A_0)$ and also assumed that 
the mass of gluino is larger than 400 GeV. 
The CP phase $\phi_t$ leads to large enhancement of the relative
mass splittings between $\tilde{t}_2$ and $\tilde{b}_L(\tilde{t}_1)$,
which in turn reduces the one-loop contribution of the stop and sbottom to 
the $\rho$ parameter. For small $\tan \beta$, such a CP violating effect
is prominent.
We have also studied how much the two-loop gluon and gluino contributions
are affected by the CP phase. Possible contributions to the $\rho$
parameter arising
from the Higgs sector with CP violation have been discussed. 

The work of JDK is supported in part by the Korea Institute for Advanced Study.

\begin{figure}
\begin{center}
\mbox{\epsfig{figure=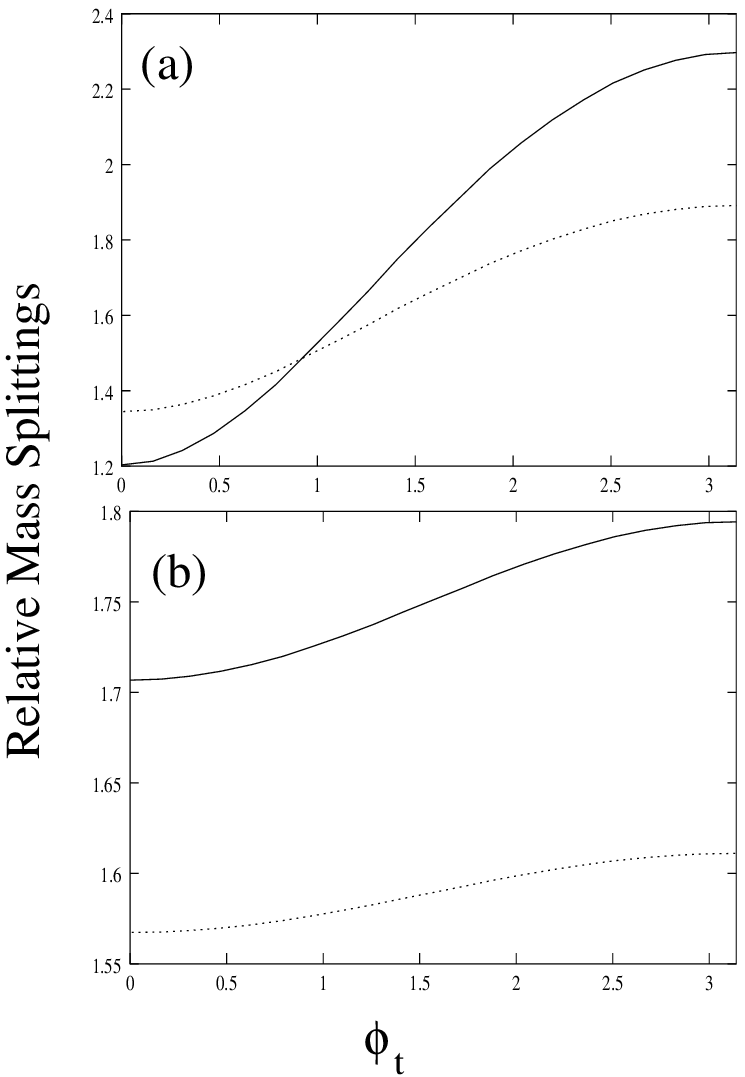}}
\end{center}
\caption{
The relative mass splittings between the two squark mass
eigenstates, $|m^2_{\tilde{t}_1}-m^2_{\tilde{t}_2}|/m^2_{\tilde{q}}$
(solid line)
and $|m^2_{\tilde{t}_2}-m^2_{\tilde{b}_L}|/m^2_{\tilde{q}}$
(dotted line)
as a function of the phase $\phi_t$ for $\tan \beta=1.6$ (a) and
$\tan\beta=20$ (b).
We take $ m_{\tilde{q}}=200$ GeV,
$|A_0|=200$ GeV, and $|\mu|=100$ GeV.
}
\end{figure}
\begin{figure}
\begin{center}
\mbox{\epsfig{figure=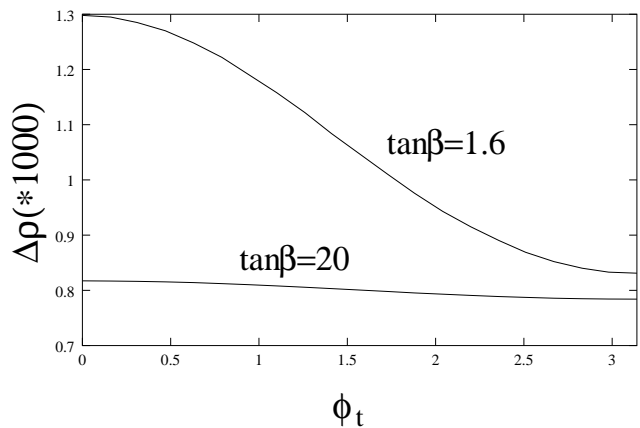}}
\end{center}
\caption{
The contribution of stop and sbottom to $\Delta\rho $ 
as a function of the phase $\phi_t$ for the same input values of the SUSY
parameters  as those in Fig. 1.
}
\end{figure}
\begin{figure}
\begin{center}
\mbox{\epsfig{figure=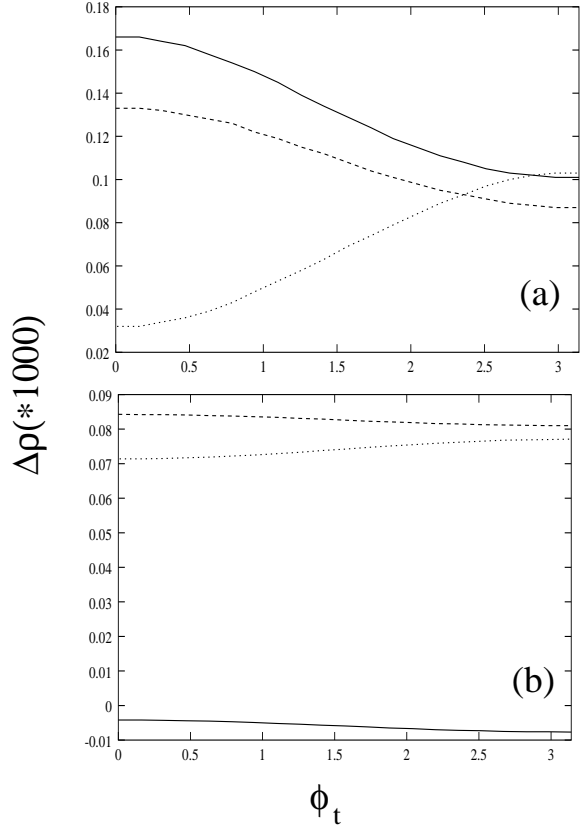}}
\end{center}
\caption{
The contributions of the two-loop QCD correction (dashed line),
the two-loop gluino (dotted line) and the Higgs boson sector (solid line)
to $\Delta\rho$  as a function of the phase $\phi_t$ for $\tan \beta=1.6$ (a)
and $\tan \beta=20$ (b).
The input values of the SUSY
parameters are taken to be the same as those in Fig. 1, and
the gluino mass and the charged Higgs boson mass to be $m_{\tilde{g}}=500$ GeV
$M_{H_c}=200$ GeV, respectively.
}
\end{figure}
\begin{figure}
\begin{center}
\mbox{\epsfig{figure=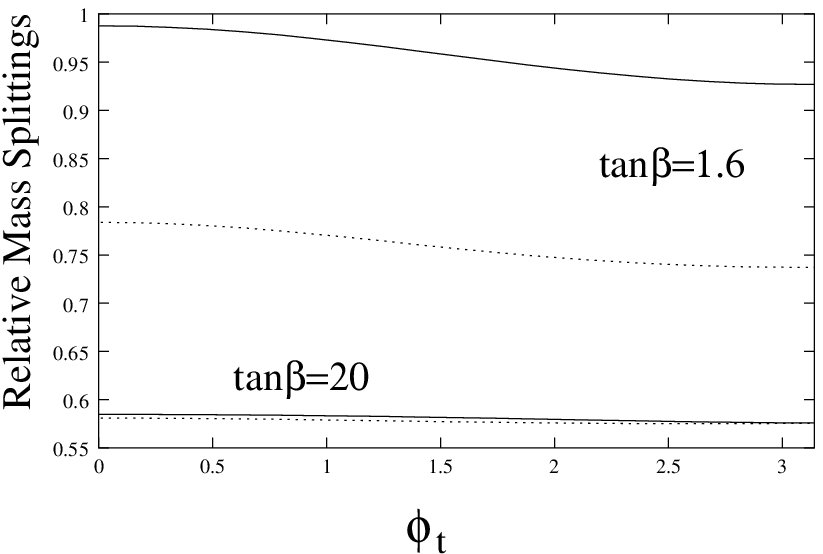}}
\end{center}
\caption{
The relative mass splittings among three neutral Higgs bosons:
$|M^2_{3}-M^2_{2}|/M^2_{H_c}$
(solid line)
and $|M^2_{2}-M^2_{1}|/M^2_{H_c}$
(dotted line).
The input values of the SUSY parameters are taken to be the same as 
the case of Fig. 3.
The lightest (heaviest) Higgs masses are denoted by $M_2(M_3)$.
}
\end{figure}

\end{document}